\documentclass[12pt]{article}
\usepackage{graphicx}
\usepackage{amsfonts}
\usepackage{amsmath}
\usepackage{amssymb}
\usepackage{hyperref}
\usepackage[brazilian]{babel}
\usepackage{indentfirst}
\usepackage[usenames,dvipsnames]{pstricks}
\usepackage{epsfig}
\usepackage{pst-grad} 
\usepackage{pst-plot} 
\title{Adapting the Euler-Lagrange equation to study one-dimensional motions under the action of a constant force}
\author{Clenilda F Dias$^1$, Vagson L Carvalho-Santos$^{2}$\\$^1$ \small Universidade do Estado da Bahia - Campus VII - Senhor do Bonfim\\$^2$ \small Instituto Federal de Educa\c c\~ao, Ci\^encia e Tecnologia Baiano \\\small Campus Senhor do Bonfim, Bahia, Brazil}
\date{}

\begin{document}
\maketitle
\begin{abstract}The Euler-Lagrange equations (EL) are very important in the theoretical description of several physical systems. In this work we have used a simplified form of EL to study one-dimensional motions under the action of a constant force. From using the definition of partial derivative, we have proposed two operators, here called \textit{mean delta operators}, which may be used to solve the EL in a simplest way. We have applied this simplification to solve three simple mechanical problems under the action of the gravitational field: a free fall body, the Atwood's machine and the inclined plan. The proposed simplification can be used to introducing the lagrangian formalism to teach classical mechanics in introductory physics courses.\end{abstract}

\section{Introduction}
In introductory physics courses, the student is presented, in general, to the newtonian formalism to describe the dynamics of a rigid body. In this formalism, the concept of force is used and the three Newton's laws depict all the relevant characteristics to understand the motion of the body. This fact requires students to think in force before to progress to concepts such as energy, momentum and least action principle. Besides, introducing the newtonian formalism requires the student to deal with the vectorial character of force, which must bring difficulties for problems where vector decomposition is demanded. In this context, the concept of force has received several criticisms. For example, Wilczec \cite{wilckzec-essay} has argued that the force, given by the Newton's second law, has no independent meaning. In addition, Jammer \cite{jammer-preface} has suggested that the concept of force is in the end its life cycle. 

The main alternative to describe the motion of a rigid body without recourse to the concept of force is the lagrangian formulation to classical mechanics, which describe the dynamical evolution of a mechanical system from the concept of least action principle. The motion equations obtained from the least action principle \cite{greiner} agree with the second Newton's law (without the necessity of the concept of force). The lagrangian formalism is not important only in classical mechanics problems, but also in several areas of physics such as quantum field theory \cite{ryder} and condensed matter field theory \cite{Atland-book}. Besides, the lagrangian formulation for field theory is behind the Noether's theorem, which connects symmetries and conservational laws \cite{ryder}. In this context, the sooner the student makes contact with this formalism, the sooner he will be able to study and discuss several relevant and current topics in physics.

Despite its importance, no attention is devoted to the lagrangian formalism for classical mechanics in the beginning of undergraduate courses of engineering, mathematics or physics. In fact, it is not a simple task to introduce this theme without speaking on the concepts of partial derivative and least action principle, which requires a mathematical knowledge that an ordinary student in introductory courses did not reach yet. However, Curtis has called attention to the importance of a qualitative discussion of modern physics themes, including the least action principle, in introductory courses of physics \cite{curtis-EJP}. Indeed, several works have been devoted to provide the student with information on relevant topics and relatively advanced in Physics \cite{curtis-EJP}. For example, Organtini \cite{organtini} and Cid \cite{cid} discuss the possibility to introduce the Higgs mechanism to undergraduate students. Besides, the study about surface plasmon resonance, which is often confusing for undergraduate students, was proposed from the link between classical concepts of resonance and the solution of problems \cite{domini}. In addition, Bezerra \textit{et al} has proposed that the introduction of the theme of magnetic dipoles can be given from a Taylor expansion of the Biot–Savart law to obtain, explicitly, the dominant contribution of the magnetic field at distant points, identifying the magnetic dipole moment of the distribution \cite{rio}.

In this context, we propose the development of a simplification of the EL to analyze the dynamics of particles under the action of a constant force, in order to obtain a simple description of a mechanical system, and using a mathematical formalism that is accessible to an undergraduate student in the beginning of a physics course. Despite the criticisms on the 'overmathematicalization' in physics education and the proposition that math must be the last thing to be taught, Taber arguments that the mathematical concepts are indispensable in physics courses \cite{taber-PEd}. Thus, we believe that this work must bring new contributions and ways to introduce modern topics of physics in introductory courses.

This work is divided as follows: in section \ref{LagrangeSup} we perform a brief review about the Lagrange formalism for classical mechanics; section \ref{LagrangeMed} presents the simplification of the EL and three simple examples are developed in section \ref{Examples}. Finally, section \ref{Conc} brings the conclusions and prospects.

\section{The Euler-Lagrange equations}\label{LagrangeSup}
In the lagrangian formalism, the mechanical system is described by $N$ generalized coordinates and $N$ generalized velocities. The system evolves from a configuration in the time $t_1$ to another in the time $t_2$. The ask to be answered is: How does the system evolve from the configuration 1 for the configuration 2? In the newtonian formalism, this evolution is given by the Newton's second law. In the lagrangian one, the Hamilton's principle (also called least action principle) describes the dynamical evolution of the system. The least action principle states that the evolution of the system from configuration 1 to configuration 2 is such that the action is a minimum.

The concept of action is associated with the lagrangian $L$ of the system, which is a function of the generalized coordinates ($q$), generalized velocities ($\dot{q}$) and time ($t$), that is, $L\equiv L(q,\dot{q},t)$. The action is defined as
\begin{equation}
S=\int_{t_1}^{t_2}L(q,\dot{q},t)dt.
\end{equation}

From the Hamilton's principle, we have that $\delta S=0$, and then
\begin{equation}
\delta S=\int_{t_1}^{t_2}\sum_{i=1}^{N}\left(\frac{\partial L}{\partial q_i}-\frac{d}{dt}\frac{\partial L}{\partial \dot{q}_i}\right)\delta q_i(t)dt=0.
\end{equation}
Once the $\delta q_i$ are arbitrary functions of $t$, if there is no links between the $q_i$'s, they will be independent and
\begin{equation}\label{EQL}
\frac{\partial L}{\partial q}-\frac{\partial}{\partial t}\left(\frac{\partial L}{\partial\dot{q}}\right)=0,
\end{equation}
where we have omitted the $i$ sub index. This is the EL, which gives us the time evolution of the dynamical properties of the system. This equation is as important to the lagrangian formalism as the Newton's second law is to the newtonian one. In fact, if the lagrangian is defined as $L=T-V$, where $T$ is the kinetic energy and $V$ is the potential energy, the obtained EL is equivalent to the Newton's second law. For proofing it in the case of conservative forces, we will consider a particle in a region where there is an interaction potential
\begin{equation}
L=\frac{1}{2}mv^2-V(\mathbf{r}).
\end{equation}
By assuming that the generalized coordinates are the cartesian coordinates, we have that 
$$\frac{\partial L}{\partial x_i}=\frac{\partial V}{\partial x_i},$$
\begin{equation}
\frac{d}{dt}\frac{\partial L}{\partial\dot{x}_i}=m\ddot{x}_i.
\end{equation}
Thus, the EL is evaluated as
\begin{equation}
m\ddot{x}_i=-\frac{\partial V}{\partial x_i},
\end{equation}
which, as expected, is the Newton's second law for conservative forces. Now, we are in conditions to simplify the EL for one-dimensional motions under the action of a constant force and study it for some particular problems.

\section{Simplification of the EL}\label{LagrangeMed}
From Eq. (\ref{EQL}), one can note that the resolution of the EL equation for a particular mechanical system is obtained from two partial derivatives, which are not a simple task for most students in the beginning of undergraduate courses. In this context, a simplification of Eq. (\ref{EQL}) can be used to introduce the lagrangian formalism in introductory courses. In order to simplify the EL for one-dimensional problems, without lost of generality, we will adopt the $x$-axis as the direction of the motion and will start from the concept of partial derivative.
 
Given an arbitrary function $f(x,y)$, the partial derivatives of $f$ in relation to $x$ and $y$ at the point $(x_0,y_0)$ are defined, respectively, by
\begin{equation}
\frac{\partial f}{\partial x}(x,y)=\lim_{\Delta x\rightarrow 0}\frac{f(x+\Delta x,y)-f(x,y)}{\Delta x}
\end{equation}
and
\begin{equation}
\frac{\partial f}{\partial y}(x,y)=\lim_{\Delta y\rightarrow 0}\frac{f(x,y+\Delta y)-f(x,y)}{\Delta y},
\end{equation}
where $\Delta x=x-x_0$ and $\Delta y=y-y_0$.

In this way, from we have that EL can be written as
\begin{equation}
\frac{\partial L}{\partial v}=\lim_{v\rightarrow v_0}\frac{L(x_0,v)-L(x_0,v_0)}{v-v_0}
\end{equation}
and
\begin{equation}
\frac{\partial L}{\partial x}=\lim_{x\rightarrow x_0}\frac{L(x,v_0)-L(x_0,v_0)}{x-x_0},
\end{equation}
which are the partial derivative definitions for the two terms of the EL.

Now, for one-dimensional motions in a constant force, we will define two operators, $\Delta_x$ and $\Delta_v$, here called \textit{mean delta operators}, which act at the lagrangian in the below form
\begin{itemize}
\item the $\Delta_x$ operator acts in the lagrangian terms that depend only on the $x$ coordinates. That is
\begin{equation}
\Delta_x L=L(x)-L(x_0);
\end{equation} 
\item the $\Delta_v$ operator, which acts in the lagrangian terms that depends only on $v$. That is
\begin{equation}
\Delta_v L=L(v)-L(v_0),
\end{equation} 
\end{itemize}
where $L(x)$ and $L(v)$ are the terms in the lagrangian depending only on $x$ and $v$, respectively. Thus, we have that
\begin{equation}
\frac{\Delta_{v}L}{\Delta v}=\frac{L(v)-L(v_0)}{v-v_0}\label{EQ3}
\end{equation} and
\begin{equation}
\frac{\Delta_{x}L}{\Delta x}=\frac{L(x)-L(x_0)}{x-x_0}\label{EQ4}
\end{equation}

Now, as well as we call the mean velocity of a particle as $v_m=\Delta x/\Delta t$, we will call the Eqs. (\ref{EQ3}) and (\ref{EQ4}) as the \textit{mean lagrangian} in relation to $v$ and $x$, respectively. With that definitions, we will rewrite the Eq. (\ref{EQL}), for one-dimensional motions subject to a constant force, as
\begin{equation}
\frac{\Delta}{\Delta t}\left(\frac{\Delta_{v}L}{\Delta v}\right)-\frac{\Delta_{x}L}{\Delta x }=0,\label{zen0}
\end{equation} 
where $\Delta x=x-x_{0}$, $\Delta v=v-v_{0}$ and $\Delta t=t-t_0$. From Eq. (\ref{zen0}) we can conclude that the temporal variation of the \textit{mean lagrangian} with relation to $v$ is equal to the variation of the \textit{mean lagrangian} in relation to $x$. Here, the Eq. (\ref{zen0}) is the simplification of the EL for one-dimensional motions subject to a constant force, without use of the concept of partial derivatives.

After the presentation of the simplification of the EL equation, we can study three simple examples for testing the validity of this approximation in which we have particles  subject to a gravitational field: free fall body, the Atwood's machine and the inclined plan.

\section{Three simple examples}\label{Examples}
\subsection{Free fall body}
A free fall body consists in a particle having mass $m$ and subject only to a gravitational field $g$. In this case, the lagrangian is given by
\begin{equation}
L=\frac{1}{2}mv^{2}-mgy,
\end{equation}
where $y$ is the height in relation to the ground. In this case, the lagrangian components depending on the position and on the velocity are given by $L(v)=\frac{1}{2}mv^{2}$ and $L(y)=-mgy$. Obviously, $L(v_{0})=\frac{1}{2}mv_{0}^{2}$ and $L(y_{0})=-mgy_{0},$ where $y_0$ and $v_0$ are respectively the initial height and initial velocity of the particle. Thus
\begin{equation}\label{anterior}
\frac{\Delta_vL}{\Delta v}=\frac{\frac{1}{2}mv^{2}-\frac{1}{2}mv_0^{2}}{v-v_{0}}=\frac{1}{2}m(v+v_0)
\end{equation}
and
\begin{equation}
\frac{\Delta_yL}{\Delta y}=\frac{-mgy+mgy_0}{y-y_0}=-mg,
\end{equation}
where we have used the factorization property $(v^{2}-v_{0}^2)=(v+v_{0})(v-v_{0})$ in Eq. (\ref{anterior}). 

Aiming to continue our analysis, we will define ${(v+v_0)}/{2}\equiv v_{_M}$, 
where $v_{_M}$ is the average of the velocities in two arbitrary points along the trajectory described during the motion of the body. For uniformly varied rectilinear motions the average of velocities is equal to the mean velocity of the particle. In order to prove it, we will start from the definition of mean velocity, given by $\text{v}_m=\Delta x/\Delta t$. In the case of a particle subject to a constant force, the motion is uniformly varied and the Torricelli's equation leads to
\begin{equation}
\Delta x=\frac{(v+v_0)(v-v_0)}{2a}=\frac{(v+v_0)\Delta t}{2}\Rightarrow \text{v}_m=\frac{v+v_0}{2}=v_{_M}.
\end{equation} 
An then
\begin{equation}
\frac{\Delta}{\Delta t}\left(\frac{\Delta_{v}L}{\Delta v}\right)=m\frac{\Delta v_{_M}}{\Delta t}.
\end{equation}

Note that in the above equation, we are not considering the variation of the mean velocity of a particle, but the variation of the average of velocities between two arbitrary points along the particle trajectory, which changes if we consider two different excerpts during the body's motion. For example, when we compare the average of velocities during the first and second halves of the motion of a free fall body, we note that they have different values​​. In fact, if some body is dropped from a height $h$, in a place where the gravitational field is $g$, when it reaches the height ${h}/{2}$, its speed is $ v_1=\sqrt{gh} $. Once it was abandoned, we have $v_0=0$ and in the first half of the fall, this average is
\begin{equation}
v_{_{M1}}=\frac{v_1+v_0}{2}=\frac{\sqrt{gh}}{2}.
\end{equation}
However, when it reaches the ground, the body velocity is $v_2=\sqrt{2gh}$, in such way that during the second half of the trajectory, the average between the velocities of the free fall body is
\begin{equation}
v_{_{M2}}=\frac{v_2+v_1}{2}=\frac{\sqrt{3gh}}{2}.
\end{equation}
Then, there exists a variation of the average of velocities of the particle along the trajectory. Finally, from the definition of mean acceleration, we have that $a\equiv({\Delta v_{_M}}/{\Delta t})$, and from Eq. (\ref{zen0}), we have $a=-g$.

Since the gravitational field is constant next to the Earth surface, the particle realizes an uniformly accelerated rectilinear motion, with velocity and position given by
\begin{equation}\label{Eq-mot}
v=v_0-gt,\hspace{1cm}y=y_0+v_{0}t-\frac{1}{2}gt^2,
\end{equation}
which are the motion equations to a particle moving under the Newton's laws, with constant force $\mathbf{P}=-mg\hat{x}$.

\subsection{The Atwood's machine}
The Atwood's machine is a simple problem that can be used to introduce the students to problems involving ropes and chains. Despite it is an ancient problem, it will be considered here by pedagogical reasons because it is of easy physical interpretation and it is useful to study mass variable systems \cite{sousa-PEd}.

\begin{center}
\begin{figure}
\includegraphics[scale=0.5]{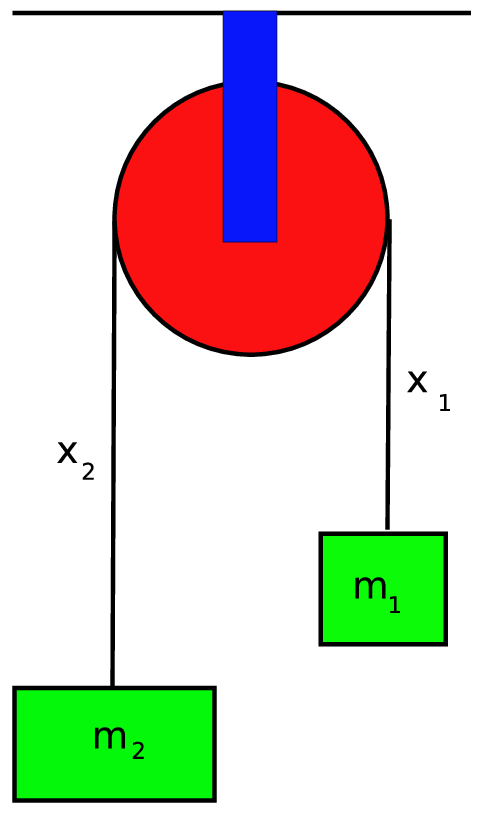}\caption{[Color online] Atwood's machine scheme. Two blocks with mass $m_1$ and $m_2$ connected by a string with lenght $\ell=x_1+x_2+\pi R$.}\label{atwoodFig}
\end{figure}
\end{center}

The Atwood's machine consists in two blocks of masses $m_1$ and $m_2$, which are connected by a massless string with length $\ell$ passing over a frictionless pulley of negligible mass and radius $R$. By taking $m_1>m_2$ and $\ell=\pi R+x_1+x_2$ (see Fig. \ref{atwoodFig}), we obtain
\begin{equation}
L(v_{1})=\frac{1}{2}m_1v_1^{2}\,\,\,\,\,\,\,\, e\,\,\,\,\,\,\,\, L(x_1)=-m_1gx
\end{equation}
and
\begin{equation}
L(v_{2})=\frac{1}{2}m_{2}v_2^{2}\,\,\,\,\,\,\,\, e\,\,\,\,\,\,\,\, L(x_{2})=-m_{2}g(\ell-x),
\end{equation}
where we have used the fact that when the block 1 downs a distance $x$, the block 2 rises $\ell-x$. Since the blocks are linked by a string with constant length $\ell$, their velocities $v_1$ and $v_2$ are equal during all the motion, that is, $v_1=v_2\equiv v$. Thus, the total lagrangian of the system is
\begin{equation}
L=\frac{1}{2}(m_1+m_2)v^2-m_1gx-m_2g(\ell-x).
\end{equation}
At the time $t_0$, we have:
 \begin{equation}
 L_0=\frac{1}{2}(m_1+m_2)v_0^{2}-m_1gx_0-m_2g(\ell-x_0).
 \end{equation}

The acceleration of the system can be obtained from using Eq. (\ref{zen0}). In this case, we will have that
  $$\frac{\Delta_{v}L}{\Delta v}=\frac{\frac{1}{2}(m_1+m_2)v^2 -\frac{1}{2}(m_1+m_2)v_0^2}{v-v_0}$$
\begin{equation}=\frac{\frac{1}{2}(m_1+m_2)(v^2-v_0^2)}{v-v_0}=\frac{1}{2}(m_1+m_2)(v+v_0)\label{zen1}
\end{equation}
and
\begin{equation}\frac{\Delta_{x}L}{\Delta x}=\frac{(-m_1gx-m_2g(\ell-x))+(m_1gx_0+m_2g(\ell-x_0))}{x-x_0}=g(m_2-m_1).\label{zen2}
  \end{equation}

By substituting Eqs. (\ref{zen1}) and (\ref{zen2}) in Eq. (\ref{zen0}), we obtain
\begin{equation}
(m_1+m_2)\frac{\Delta v_{_M}}{\Delta t}-g(m_1-m_2)=0\Rightarrow a=\left(\frac{m_2-m_1}{m_2+m_1}\right)g.
  \end{equation}
Then, as expected, we have found the acceleration predicted by the Newton's law \cite{sousa-PEd}. Once the acceleration is a constant, the motion equations are given by Eq. (\ref{Eq-mot}) with the replacement of $g$ by $g({m_2-m_1})/({m_2+m_1})$.

\subsection{The inclined plan}
\begin{figure}
\includegraphics[scale=0.8]{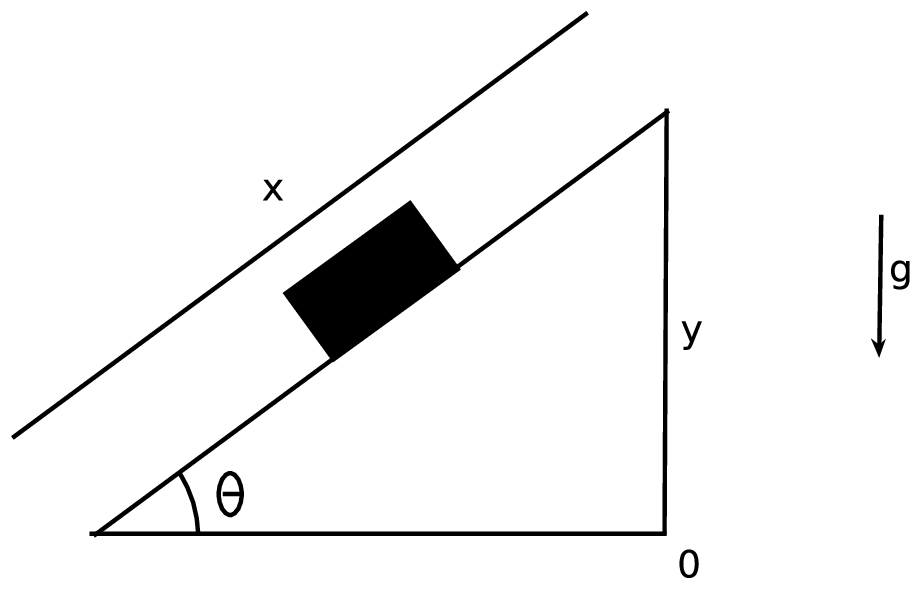}
\caption{Inclined plan scheme. The body slips on an inclined plan by an angle $\theta$.}\label{incplan}
\end{figure}

The last example to be treated in this paper is the inclined plan, which consists in a block with mass $m$ on a plan inclined by an angle $\theta$, in the presence of a gravitational field $g$ (See Fig. \ref{incplan}). Supposing that the block is left from the rest, we have that
\begin{equation}
L=\frac{1}{2}mv^2-mg{x}{\sin\theta},
\end{equation}
where $x$ is the distance measured on the surface plan ($x=0$ in the bottom of the plan). In this case, we have that
\begin{equation}\frac{\Delta_{v}L}{\Delta v}=\frac{\frac{1}{2}mv^2}{v}=mv_{_M}\end{equation}
and,
\begin{equation}
\frac{\Delta_{x}L}{\Delta x}=\frac{-mgx\sin\theta}{x}=-mg\sin\theta.
\end{equation}
Finally, from Eq. (\ref{zen0}), the acceleration is evaluated as
\begin{equation}
a=-g\sin\theta,
\end{equation}
which is the acceleration predicted by the Newton's law, as it should be. Once the angle is not variable, the acceleration of the block is constant, and if the friction in the contact between the plan and block surfaces is negligible, the block slips on the plan performing an uniformly variable rectilinear motion.
It is important to note that we have obtained this result without the use of the concept of forces and without the need of vectors decomposition.

\section{Conclusions and prospects}\label{Conc}
In this work, we have proposed that the lagrangian formalism can be presented in introductory physics courses for undergraduate students. In this context, from introducing two operators, here called \textit{mean delta operators}, we have simplified the EL in such way that it may be solved by students in introductory physics courses even then they do not know mathematical techniques to solve partial derivatives. We have used the described simplification to solve three examples: the free fall body, the Atwood's machine and the inclined plan. In the three cases, the obtained motion equations agree with that predicted by the Newton's second law, as it should be.

The proposed simplification has the limitation to be applied only to one-dimensional problems in which the particle is under the action of a constant force. However, this work opens new possibilities in the discussions about the removal of the force concept from mechanical courses, once this one is in the end of its life cycle \cite{jammer-preface}. Among the perspectives that can be opened up with this work, there is the possibility to extend this formulation for more complex problems.\\

\begin{center}
\textbf{Acknowledgements}
\end{center}
The authors thank CNPq (Grants number 401132/2016-1 and 301015/2015-5) for financial support. We are also grateful to Ivan S Costa and Maria Aparecida (Cidinha) by encouraging the development of this work. V L Carvalho-Santos thanks the hospitality of the Universidade do Estado da Bahia - Campus VII, where part of this work was developed.\\

\end{document}